\documentstyle[astro,dinA4,psfig,11pt]{article}
\begin{document}
\title{Jets in AGN -- New Results from HST and VLA}
\author{Heino Falcke}
\date{Max-Planck-Institut f\"ur Radioastronomie, Auf dem H\"ugel 69, D-53121 Bonn, Germany
(hfalcke\atsign{}mpifr-bonn.mpg.de)}

\maketitle

\centerline{\it to appear in: Reviews in Modern Astronomy, Vol. 11}
\centerline{\it R.E. Schielicke (Ed.), Astronomische Gesellschaft 1998}

\begin{abstract} This paper summarizes some of our recent projects
which try to illuminate the nature and importance of jets associated
with active nuclei and compact objects. After a short introduction on
jets in radio galaxies and radio loud quasars the paper focuses on
jets in other source types, such as radio-quiet quasars, Seyfert and
LINER galaxies, and stellar mass black holes. Radio observations of
quasars, for example, have brought new evidence for the existence of
relativistic jets in radio quiet quasars, while HST and VLA
observations of Seyfert galaxies have now clearly established not only
the presence of radio jets, but also the great importance these jets
have for the morphology and the excitation of the
emission line region in these AGN. Moreover, a recent VLA survey found
a large fraction of low-luminosity AGN to host compact, flat-spectrum
radio cores indicating the presence of radio jets there as
well. Finally the jet/disk-symbiosis model, which successfully
explains radio cores in LINER galaxies, is applied to the stellar mass
black hole GRS~1915+105, indicating that the radio cores in both types
of sources are just different sides of the same coin. The conclusion
drawn from all these observations is that radio jets are a ubiquitous
feature of most---if not all---AGN and play an important effect in
the overall energy budget, as well as for the interpretation of
observations in other wavebands (e.g. optical emission lines).
\end{abstract}

\section{Introduction} 
One of the main subjects for radio astronomers has been the study of
extragalactic radio jets. When observed at a higher resolution many of
the first radio sources which were discovered in the early years of
radio astronomy later turned out to be powerful, collimated plasma
flows of relativistic plasma (jets) which were ejected from the
nucleus of giant elliptical galaxies. These structures can reach sizes
of several million light years and hence extend way beyond their host
galaxies into the vastness of intergalactic space. This relative
isolation is ideal to study the physics of astrophysical plasma flows
in great detail (see e.g.~Bridle \& Perley 1984, Bridle et al.~1994,
Marti et al.~1997) and allows to make some estimates of the properties
of the IGM (inter-galactic medium, e.g. Subrahmanyan
\& Saripalli 1993). An even more important aspect of radio jets,
however, is that they are the most visible sign, the "smoking gun" so
to speak, of Active Galactic Nuclei (AGN) which are thought to be
powered by accreting black holes. Hence, jets have been studied with
great interest over many years and a huge zoo of different jet species
has emerged which will be discussed as a useful background for the
further discussion in the next section.

\section{Jet zoology}
The main characteristics of radio jets, are the size (compact,
i.e. parsec to kiloparsec scale, or extended, i.e. tens to hundreds of
kiloparsecs), and the spectral index (flat or steep) of the radio
sources. Steep radio spectra ($\alpha<-0.5$,
$S_\nu\propto\nu^{\alpha}$) are due to optically thin synchrotron
emission from large, extended radio sources, while flat radio spectra
can be produced if the source is very compact and the spectrum is
dominated by radiation from a number of optically thick synchrotron
components.

A typical powerful radio galaxy---a so called Fanaroff \& Riley
(1974) type II radio galaxy (short FR\,II)---consists
of two steep-spectrum, extended lobes connected by very faint (if
visible at all), well-collimated plasma beams, and a compact
flat-spectrum core (which will be discussed in more detail in
Sec.~5-7). FR\,II jets are produced by the most powerful AGN,
 have the largest kinetic powers of any jets observed
(i.e. Rawlings \& Saunders 1991), and
reach the very largest sizes observed. Probably due to their huge
powers FR\,II jets can plow with relativistic speeds through the
ambient medium, self-shielded by a huge cocoon (Begelman \& Cioffi
1989), until they slow down and terminate in a huge shock---the hot
spots. Behind the shock the material disperses and at least a part of
it flows back toward the galaxy it was ejected from. This can be seen
on many of the beautiful high dynamic-range VLA maps made in recent years
(e.g. Black et al.~1992, Leahy et al.~1997).

Such beautiful structures, however, are not the rule but rather the
exception: at lower powers (and low power AGN are naturally more
frequent than high power AGN) the jet morphology seems to change and
FR\,II radio galaxies suddenly turn into FR\,Is, where instead of a
well collimated pencil-like beam with a well-defined terminus, the jet
has a larger opening angle, entrains material from the ISM (DeYoung
1993, Bicknell 1994), becomes bright, and then slowly fades along the
way. Clearly, in those cases the jet is no longer independent of its
environment but interacts with the ISM of the host galaxy. Initially
we were only able to see the effects of this interaction on the radio
jet itself, but now we can also see the other side of this coin in
x-ray observations, which indicate how the radio plasma pushes against
gas in the galaxy (B\"ohringer et al.~1993; Holloway et al.~1996;
Clark et al.~1997).

Not always, however, are we so fortunate to see all details of the
extended jets. If, for example, those jets happen to be seen under a
very small aspect angle with respect to the line of sight,
relativistic effects will become very important. Since the jet has
velocities close to the speed of light in the nucleus, the emission
from the flat-spectrum radio core will be boosted by the relativistic
Doppler effect and for small inclination angles will become so bright
that it dominates the entire radio emission, overwhelming even the
bright extended lobes (even though they still can be found in
high-dynamic range observations, e.g. Kollgaard et al.~1992). These
galaxies appear as compact, core-dominated flat-spectrum radio
galaxies, sometimes called Blazars. If observed at very
high-resolution Blazars often show superluminal motion (an optical
illusion caused by the relativistic motion of bright features in the
jet) which is accompanied by strong flux variability down to scales of
less than a day (Wagner \& Witzel 1995; Zensus 1997). In addition one
finds luminous high-energy emission, such as x-ray and gamma-emission
(up to 10 TeV, e.g. Zweerink et al.~1997), most certainly produced by
scattering processes (e-$\gamma$, p-$\gamma$, or p-p) from
relativistic particles within the jet (Mannheim \& Biermann 1992;
Dermer \& Schlickeiser 1993).  In fact, for some souces most of the
observed luminosity is seen in gamma-rays.

But not all compact radio galaxies are Blazars, some galaxies show
steep radio spectra (at least at high frequencies), yet, instead of
penetrating deep into the inter-galactic medium (IGM) as FR\,I and
FR\,IIs do, they get stuck inside their host galaxies, either because
they are frustrated or simply too young. These sources are called CSS
(compact steep-spectrum) or GPS (Gigahertz-peaked spectrum) sources
(see O'Dea 1998 for a recent review). Looked at with higher-resolution
one can resolve the jets and find extended lobes on scales of
several kpc (for CSS) down to hundreds of parsecs (GPS, needs VLBI)
--- at least some of them must be the predecessors of the large FR\,I
and FR\,II radio galaxies.

The main reason why these galaxies have been studied in such detail
so far is their large radio fluxes of 100mJy up to several tens of
Jansky, which makes them easily accessible with current
technology. Hence, when we discuss the properties of relativistic jets
in AGN, we usually tend to think exclusively about those radio
galaxies, radio-loud quasars, and Blazars.  But is this the whole
universe, or just the tip of the iceberg? In the following I will
discuss a number of other classes of AGN where jets have become or
will become an important issue.

\section{Jets in `radio-weak' AGN}
In comparison to stellar winds it is often argued that the escape
speed from the central object is an important factor that determines
the terminal jet speed. If that is true and since we believe that most
of the AGN are powered by a black hole one should assume that if an
AGN produces a jet it should {\it always} be
relativistic. Consequently the crucial question then becomes: Do all
AGN have jets? In Falcke (1994) and Falcke \& Biermann (1995) we
proposed that, since black holes do not have many free parameters, AGN
should be similar in their basic properties (``the universal engine'',
Falcke 1996a) and hence one should {\it ab initio} assume that all
AGN, rather than only a few sub-classes, have relativistic jets. Using
Occam's razor we also suggested that jets and accretion process
(accretion disk) should form a symbiotic system in the sense that both
are always required for an AGN. As it turned out, this hypothesis, in
its simplicity, was surprisingly successful and has motivated most of
the research presented here. In a review Livio (1997) comes to a
similar conclusion, i.e. that a majority of accretion disk systems
produce jet-like outflows, however, negates the necessity of jets for
accretion disks by hinting at CVs where jets are not
observed. Fortunately in a later paper (Shahbaz, Livio, Southwell, \&
Charles 1997) the same author is the first to report evidence for a
jet-like outflow in a CV.

\subsection{Radio-Quiet Quasars}
One class of sources where the jet/disk-symbiosis principle was used
first was the UV/radio-correlation of quasars (Falcke, Malkan,
Biermann 1995). If one looks at the distribution of the
radio-to-optical flux ratios ($R$-parameter) of an optically selected
quasar sample (here the PG quasar sample) one finds a clear dichotomy
between radio-loud and radio-quiet sources. This is especially true if
one selects only  steep-spectrum quasars, which are supposedly
unaffected by orientation effects (see Fig.~1, top). VLA observations
of the steep-spectrum radio-loud PG quasars (Miller, Rawlings, \&
Saunders 1993) and Kellermann et al.~(1994) have clearly established,
that they have FR$\,$II-type radio jets.  The radio dichotomy was
occasionally attributed to the fact that radio-quiet quasars do not
show and do not have radio-jets at all. However, as we all know,
`absence of evidence is not evidence of absence' --- especially not,
if one has not even looked yet, and following the jet/disk-symbiosis
principle, one would rather suppose that radio-quiet quasars have jets as
well.

\subsection{Predictions for boosted radio-quiet jets}
We therefore have to ask: How can we obtain evidence for or against
the presence of jets in radio quiet quasars? One direction to go would
be to look for relativistic boosting. In an optically selected sample,
we would expect that, if radio-quiet quasars have relativistic jets,
some of these quasars are accidentally pointing towards us, thus
producing a population of `weak blazars' with the following
properties:
\begin{itemize}
\item[a)]  similar to flat-spectrum, core-dominated, variable radio
quasars but with relative low radio-to-optical flux ratio ($R$),

\item[b)] apparent brightness temperatures close to $\sim10^{12}$K or above,

\item[c)] superluminal motion,

\item[d)] very faint (i.e.~radio-quiet) extended radio emission,

\item[e)] number of sources in a well selected sample, and their
Doppler-boosting relative to radio-quiet quasars both imply the same
Lorentz factor,

\item[f)] luminosity- and $z$-distribution consistent with radio-quiet
parent population,

\item[g)] host galaxies compatible with those of radio-quiet quasars.
\end{itemize}

This list is quite helpful, as it allows an either/or decision: if we
do not find a population of weak blazars, we can exclude that
relativistic jets in radio-quiet quasars exist (or one would have to
invent an argument why these jets never point towards us); if we find
them, we can prove that radio-quiet quasars must have relativistic
jets. Interestingly, in the PG quasar sample we indeed find a
population of quasars, which at least partially fulfill most of the
criteria listed above and most likely are such weak blazars.

\subsection{Radio-intermediate quasars}
\begin{figure}[t]
\centerline{
\psfig{figure=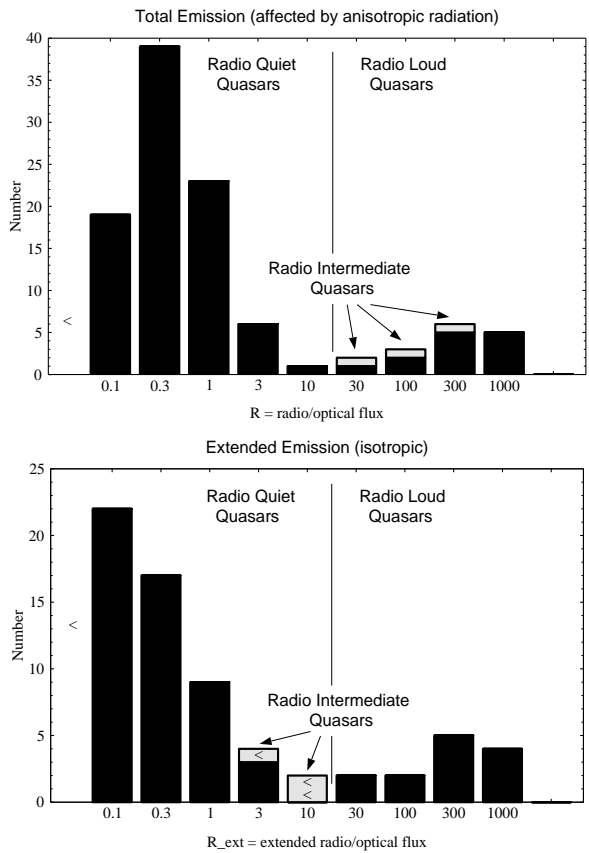,height=0.5\textheight,bbllx=4.3cm,bblly=3.2cm,bburx=18.7cm,bbury=24.1cm}
}
\caption[]{Distribution of the $R$-parameter (ratio between radio and
optical flux) for an optically selected sample of quasars with steep
radio spectra (from Falcke et al.~1996a). The RIQ are added with gray
shades. While in total flux they rival some of the fainter radio loud
quasars with their bright, flat-spectrum radio core, their extended
emission is comparable only to radio-quiet quasars.}
\end{figure}
Miller et al.~(1993) and Falcke et al.~(1995 \& 1996b) identified a
small sample of radio-intermediate quasars (RIQ) which sparsely fill
the space in $R$ between radio-loud and radio-quiet quasars. They have
optical+UV luminosities between $10^{45}$ and $10^{47}$ erg/sec, just
like the bulk of the radio-quiet quasars and unlike radio-loud quasars
which can be found only above $10^{46}$ erg/sec in the PG sample. They
are typical flat-spectrum, core-dominated quasars, but their $R$
parameter is too low for them to be boosted radio-loud quasars. If
they were boosted radio quiet quasars instead, their number and
$R$-distribution would indicate a bulk Lorentz factor of $\gamma_{\rm
j}$=2-4.  For at least the three low-redshift RIQ, there is no
extended emission above a level of a few mJy---far below what is
expected for any radio-loud quasar---neither on the VLA A- \& D-array
(Kellermann et al.~1994) nor on the EVN \& MERLIN scales (Falcke et
al. 1996a; see also Fig.~1, bottom). At least one source, III Zw~2,
has shown outbursts indicating a brightness temperature of $10^{12}$
K (Ter\"asranta \& Valtaoja 1994) which requires relativistic
boosting, while VLBI observations of the three low-$z$ sources
indicate at least lower limits of several $10^{10}$ K. III~Zw~2 is the
most interesting source in this respect since it is the most variable
source with a huge outburst every few years but which has so far
resisted attempts to resolve any structure with VLBI (e.g. Kellermann
et al.~1998).

In the meantime, since the early papers have been published, one other
prediction has been verified. In Falcke et al.~(1996a), we suggested
that in order to test the idea of the RIQ being intrinsically
radio-quiet, at least half of the flat-spectrum RIQ should have spiral
host galaxies. So far, powerful radio galaxies and radio-loud quasars
have turned out to reside in elliptical hosts, while radio-quiet
quasars seem to reside in a mix of spiral and elliptical galaxies
(Kukula et al.~1997a). Luckily, two of the three low-redshift RIQ were
part of recent host galaxy studies: HST observations of PG 1309+355
(Bahcall et al.~1997) and NIR observations of III~Zw~2 (Taylor et
al.~1996) have now shown that indeed both galaxies are spirals. This
finally confirms that the RIQ cannot be and never will be radio-loud
quasars (as they have been classified occasionally in the
past)---unless they merge and form an elliptical galaxy perhaps.

In summary, the so far identified and studied RIQ meet all the
requirements for intrinsically radio-quiet quasars, whose relativistic
jets accidentally point towards us. Moreover, since only a very small
fraction of quasars will point towards us, one can infer also that  a large
number---if not all---of the remaining radio quiet quasars must harbor
relativistic jets.

\subsection{Direct observations of jets}

The only piece missing now is direct confirmation of relativistic jets in
radio-quiet quasars; specifically superluminal motion has not yet been
observed. This is, however, not surprising given the observational
difficulties for these weak sources. VLBI observations of radio-quiet
quasars have just recently begun and even they lack the sensitivity to
detect additional components besides the core (Blundell,
priv. comm.). Deep, long-integration VLBI observations of radio-quiet
quasars are certainly needed. For example we are currently performing
a study of III~Zw~2, where we monitor the source and wait for an
outburst which then can be observed directly with the VLBA. The last
outburst in December/January 1997/98 seems to bevery promising and
preparations for the VLBI observations are under way. A second route
to follow are deep, high-resolution VLA observations of radio-quiet
quasars to look for evidence of the extended radio structures which
were seen already in some snapshot maps. First results of such a project
seem to indicate that radio quiet quasars indeed harbor Seyfert-like
jets (Kukula et al.~1997b)

Fortunately, already now are results available which can, at least in
part, answer whether direct evidence for jets in radio-quiet quasars
exists at all. First of all VLA observations of Kellermann et
al.~(1994) have already revealed a number of radio-quiet quasars with
weak, bi-polar radio-structure. Secondly, there is a large regime,
where the Seyfert galaxies and quasar classifications blend into each
other and it may be useful to study Seyferts rather than quasars,
which are closer and appear brighter on the sky, and their jets in
greater detail. In the next section I will report some results we have
obtained using the HST and the VLA for Seyfert galaxies and which
might be helpful for the interpretation of quasars as well.

\section{Seyferts}
Seyfert galaxies were first noted because of their strong emission
lines coming from their nucleus, which is emission of hot gas ionized by
an AGN. After the advent of the VLA a number of radio surveys have
shown that, besides their extended emission line regions, Seyferts also
possess -- sometimes very faint -- radio emission which is very often
bi-polar (Ulvestad \& Wilson 1984, and previous papers). Seen at
higher resolution one finds a strong tendency for the circumnuclear
emission-line and radio morphologies to be aligned in Seyfert galaxies
(e.g.~Unger et al.~1987; Pogge 1988; and Haniff, Wilson \& Ward 1988).
This strongly suggested that the ejection of the radio plasma and the
excitation of the emission line gas were related and that the ionizing
radiation escapes preferentially from the active nucleus along the
radio axis. Here the Hubble Space Telescope (HST) has made an enormous
impact: seen with the superior resolution of this telescope the
structure of the emission lines gas was revealed and in some case
shown to be well-defined cones (e.g.\ NGC~1068, Evans et al.\ 1991;
NGC~5728, Wilson et al.\ 1993; NGC~5643, Simpson et al.\ 1997), which
seemed to confirm an anisotropic escape of ionizing photons from the
nucleus. This is most popularly explained by the presence of an
optically thick `obscuring torus' (Antonucci 1993), which is able to
collimate the intrinsically isotropic ionizing radiation (see, e.g.,
Storchi-Bergmann, Mulchaey \& Wilson 1993) from the AGN. In addition,
a number of galaxies display the `ionization cone' morphology when an
excitation map is made, e.g.\ in
[\ion{O}{3}]/(H$\alpha$+[\ion{N}{2}]).

\begin{figure}[t]
\centerline{
\psfig{figure=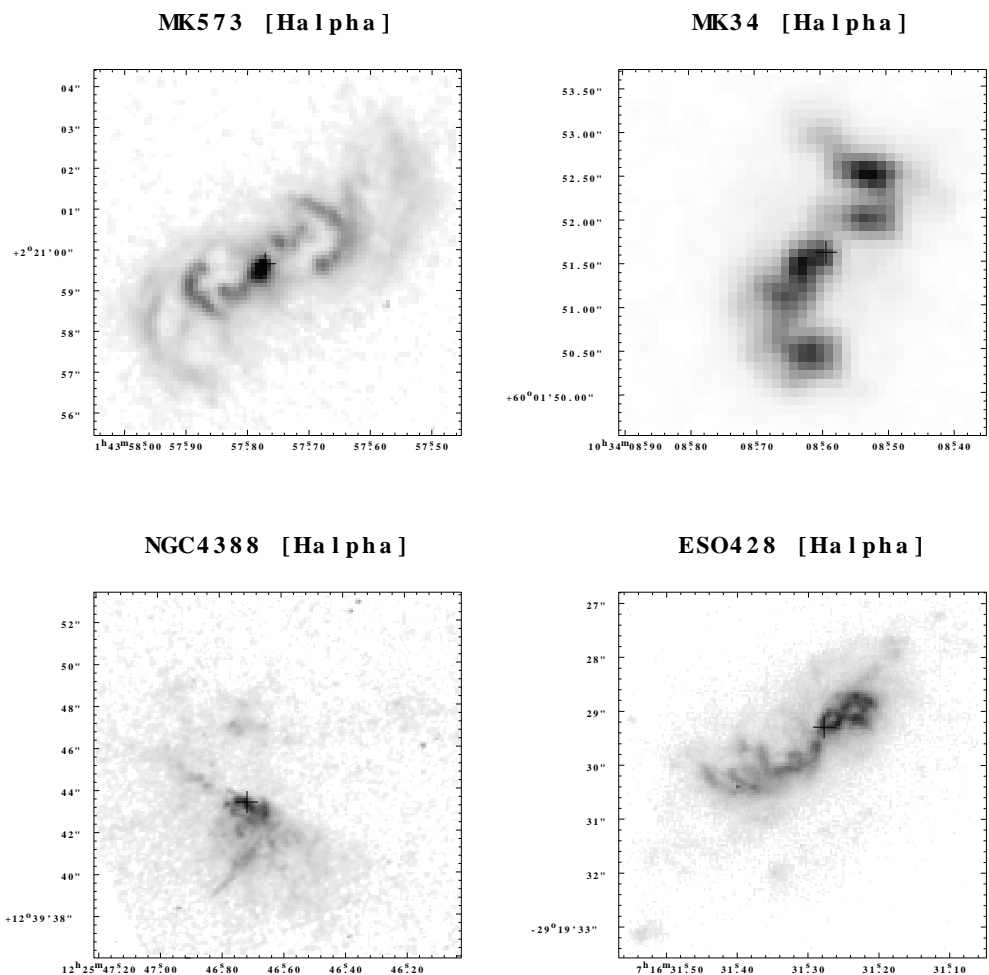,width=0.95\textwidth,bbllx=1.2cm,bblly=7.3cm,bburx=19.6cm,bbury=25.4cm}
}
\caption[]{Narrow band HST images of Seyfert 2 galaxies in the H$\alpha$ emission
line from Falcke et al.~(1998)
}
\end{figure}

The close connection between the radio ejecta of Seyfert nuclei and
their narrow line regions (NLRs) initially became apparent from their
similar spatial extents and from strong correlations between radio
luminosities and [\ion{O}{3}]$\lambda$5007 luminosity and line width
(de Bruyn \& Wilson 1978; Wilson \& Willis 1980; Whittle 1985, 1992).
Spectroscopic studies of the NLR (Baldwin, Wilson \& Whittle 1987;
Whittle et al.\ 1988), have revealed that the kinematics of the gas
are often clearly affected by the radio jets. Such interactions could
play a role in determining the structure of the NLR within the region
ionized by the nucleus. In a handful of cases, HST has shown a clear
spatial correspondence between the radio and emission-line
distributions (e.g.~NGC~5929, Bower et al.~1994; Mrk~78, Capetti et
al.~1994, 1996; Mrk~1066, Bower et al.\ 1995; Mrk~3, Capetti et al.\
1996; ESO~428--G14, Falcke et al.\ 1996c), indicating that the radio
ejecta strongly perturb the ionized gas, at least in these objects. It
has also been suggested that the hot gas associated with the shocks
generated by the interaction between the radio ejecta and the ambient
medium is a significant source of ionizing radiation (e.g.~Dopita
1995; Dopita \& Sutherland 1995; Bicknell et al.~1997; see also
reviews in Morse, Raymond, \& Wilson 1996 and Wilson 1996).

It is therefore of great importance to study more Seyfert galaxies at
the high spatial resolution which only HST can provide, to determine
whether the morphology of the narrow line region is determined by the
nuclear ionizing radiation or by the interaction of radio jets with
the interstellar medium, or by a combination of both.

\begin{figure}[t]
\centerline{
\psfig{figure=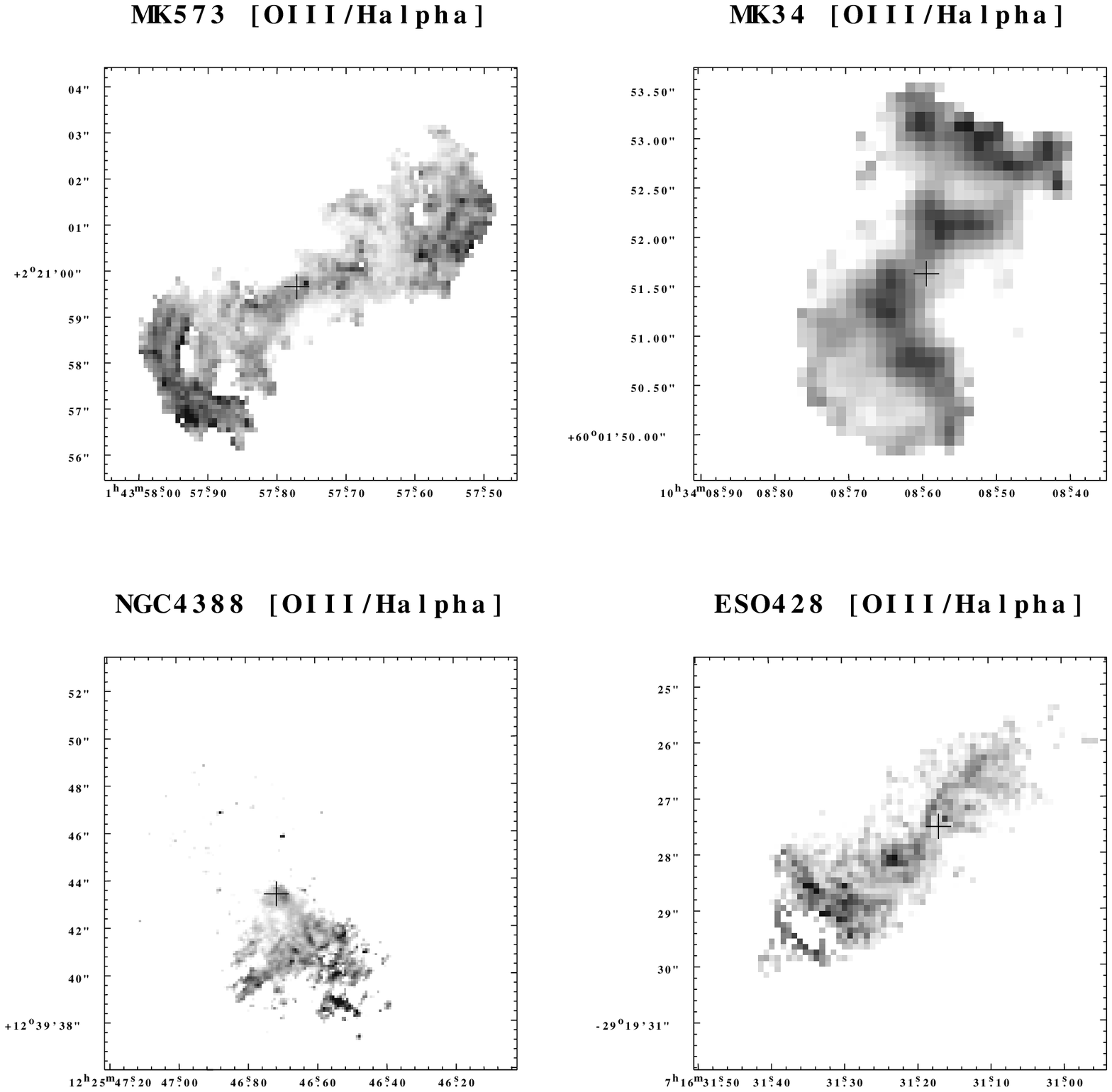,width=0.95\textwidth,bbllx=1.2cm,bblly=7.3cm,bburx=19.6cm,bbury=25.4cm}
}
\caption[]{Excitation maps (\ion{O}{3}/H$\alpha$) for the galaxies in
Fig.~1. Dark shades indicate higher excitation. The overall structure
of the highly excited gas resembles cones and bi-cones.}
\end{figure}

In Falcke, Wilson, \& Simpson (1998) we presented images taken with
the Wide Field and Planetary Camera 2 (WFPC2) of seven Seyfert~2
galaxies (see Fig.~2), selected on the basis of possessing either
extended emission-line regions (as seen in ground-based images) or
broad lines in polarized light. For each galaxy images in the light of
the [\ion{O}{3}]~$\lambda$5007 line and the
H$\alpha$+[\ion{N}{2}]~$\lambda\lambda$6548,6583 blend were taken. In
addition we also obtained new radio maps taken with the Very Large
Array (VLA), almost all in `A-configuration', providing an angular
resolution comparable with that of the HST images. Taken together,
these allowed us to compare directly the structures of the
line-emitting gas and radio plasma on scales of tens of parsecs.

And indeed in four of the seven galaxies (Mrk~573, ESO~428$-$G14,
Mrk~34, NGC~4388) we found bi-polar structures in the excitation maps
(i.e. the maps obtained by dividing the [\ion{O}{3}] by the H$\alpha$
map, Fig.~3) and a number of finer structures in the emission line
regions of all galaxies (with the exception of one, Mrk 1210, which
was basically unresolved).  In addition, the high quality radio maps
of the galaxies we obtained, show the considerable diversity one can
find in the radio structure of Seyferts, all of which indicate
the presence of a jet outflow: narrow, filamentary jets
(ESO~428$-$G14), triple structures with a core and two hotspots
(Mrk~573), jets plus two hotspots (Mrk~34), radio plumes and
limb-brightened lobes (NGC~4388), etc. (Fig.~4).

Even though the dynamic range of Seyfert radio maps is naturally lower
than what one can obtain for the much brighter radio galaxies, this
diversity is much larger than what we find in the latter. This is of
course readily understood, since Seyfert jets are much more subject to
jet-ISM interaction than FR\,II radio galaxies, because of their
orders of magnitude lower absolute jet powers. Morphologically,
this interaction can be seen in many images of the HST: e.g. in
Mrk~573 and Mrk~34 the radio hotspots coincide with regions of reduced
excitation, the bow-shock structure in the emission-line gas of
Mrk~573 is most certainly caused by the action of an outflow as
indicated by the presence of the radio hotspots, and the filamentary
emission-line structure in ESO~428--G14 finds its detailed counterpart
in the filamentary radio jet.

\begin{figure}[t]
\centerline{
\psfig{figure=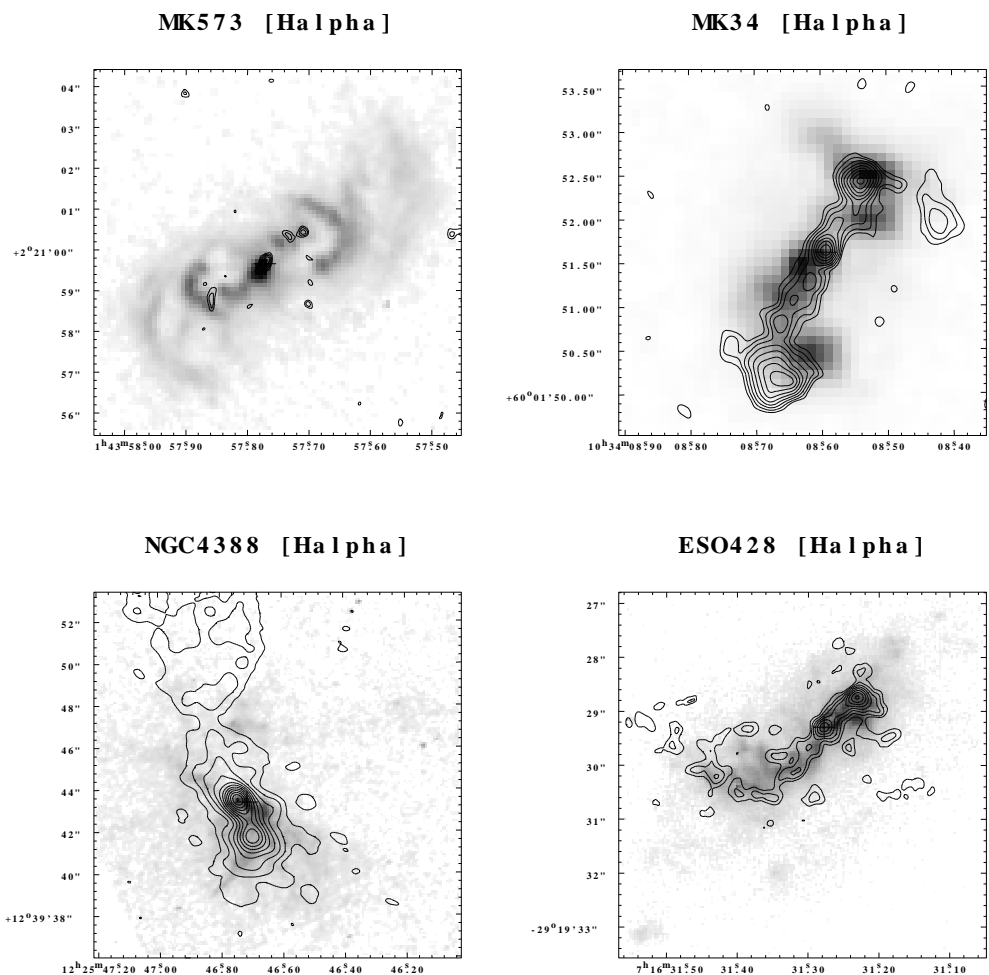,width=0.95\textwidth,bbllx=1.2cm,bblly=7.3cm,bburx=19.6cm,bbury=25.4cm}
}
\caption[]{The same images as in Fig.~1 with the VLA
radio contours overlaid. The observed wavelengths are 3.5cm for Mrk~34
and NGC~4388 and 2 cm for Mrk~573 and ESO428--G14. The direct relation
between radio and optical emission is obvious.}
\end{figure}

It is quite obvious from this data that there is not only a close
correlation between the radio and the emission-line morphologies, but
that {\it the radio jet-ISM interaction is an important effect which
strongly influences the excitation and morphology of the NLR}.
 
Does this mean the unified scheme is wrong? Is the bi-polarity of the
NLR and are the ionization cones just illusions of a weary
astronomer's soul longing for a simple scheme to explain the Seyfert
world? Are {\rm all} the structures seen in the NLR produced by
widening, self-excited matter outflows as Dopita and others suggest?

Fortunately, in at least two cases we can find good arguments, that the
anisotropic photon escape scheme remains still valid: In Mrk~573 and
NGC~4388 `ionization cones' (Pogge \& De Robertis 1995; Capetti et
al.~1996; Pogge 1988; Corbin et al.~1988) were already known from
ground-based observations on the arcsecond scale and we found an
equivalent counterpart on the sub-arcsecond scale with identical
opening angle. This continuation and the straightness of the cones
clearly favor some kind of obscuring `torus' with well defined inner
edges around the nucleus that leads to a beamed ionizing
continuum. Attributing these cones to the action of an outflow seems
unreasonable, since the radio ejecta we do see have not only a very
different appearance in both galaxies despite similar cone structures,
the jets are also themselves subject to collimation by the galaxy,
i.e. as seen in the constriction of the northern lobe of NGC~4388, and
therefore are nothing like the proposed freely expanding, conical
outflows. On the other hand, Mrk~573 has not only a straight
`excitation cone' but, with its bow-shaped emission line strands, is
also the clearest example of a jet shaped NLR. Consequently, any
successful model of the NLR of this galaxy will require a {\it composite
model that includes photo-ionization from a central source {\rm and}
the impact of a radio jet}.

Other examples for the shaping of the NLR by the jet are ESO~428$-$G14
and NGC~4388. The former exhibits well collimated, irregular
emission-line strands on one side and a figure ``eight'' morphology on
the other. The latter has been interpreted as two helical
emission-line strands wrapping around a radio jet (Falcke et
al.~1996c). The overall structure of the NLR and of the radio jet in
ESO~428--G14 is perhaps the most bizarre one can find.  In NGC~4388 a
bright spike is seen in the ionized gas at the end of the southern
jet, while the radio plasma to the north flows apparently unhampered
out of the galaxy disk and forms a large ($\sim$ 1 kpc) radio
plume. This structure is reminiscent of the radio lobe found, for
example, in NGC~3079 (Seaquist et al.~1978) and a few other galaxies
(Ford et al.~1986). This kind of limb-brightened radio lobe stands in
marked contrast to the well-collimated, stranded jets in ESO~428$-$G14
and NGC~4258 (e.g.~Cecil, Wilson, \& Tully 1992). An important
difference is that NGC~3079 and NGC~4388 have jets which escape almost
perpendicular to the galaxy plane, while in NGC~4258 and perhaps
ESO~428--G14 the jet appears to be directed into the disk of the
galaxy. The difference in radio morphology may then be ascribed to the
much higher external gas density when the jet is in the disk rather than in
the galaxy halo.

What remains unclear, however, is the exact nature of the jet/ISM
interaction in our Seyferts. To what degree are jet induced shocks
responsible for the excitation of the gas? Are there regions, where
shock excitation dominates the photo ionization from the nucleus? What
amount of energy is locally dissipated in the jets due to this
interaction? To answer these questions the observation of only two
emission lines is not enough and long-slit spectroscopy will be
needed. For ESO~428--G14, the necessary HST time for such observations
has been allocated, and other projects will address similar questions
in the future. Moreover, some recent results obtained with the FOC
on board the HST already now strongly support the jet/ISM interaction
picture (Winge et al.~1997, Axon et al.~1997).

At least in a very simple way the jets must have an influence on the
excitation of the gas: e.g. the reduced excitation (i.e.~
[\ion{O}{3}]/(H$\alpha$+[\ion{N}{2}]) ratio) of the inner
emission-line arcs in Mrk~573 is consistent with a lower ionization
parameter which can be understood if the arcs represent gas which has
passed through a radiative bow shock, cooled and increased in
density. A similar effect is seen in Mrk~34, in which the radio lobes
coincide with a low-excitation region, while the jet itself is
surrounded by high-excitation gas. The wiggles seen in the radio jet
of this galaxy could possibly be interpreted as some kind of
Kelvin-Helmholtz instabilities between the radio plasma and the
surrounding, ionized gas. Looking at all the galaxies in our sample,
there seems to be indeed a tendency for radio lobes to coincide with
lower excitation regions, presumably a result of compression of the
ambient gas.

Finally, we can now come back to our initial question about the
presence and importance of jets in general. Especially with respect to
the situation in radio-quiet quasars, we are now much more prepared to
give a positive answer. Some of our galaxies (e.g. Mrk~34 and Mrk~573)
have [\ion{O}{3}] luminosities comparable to radio-quiet, low-redshift
quasars. Hence, for such objects, we are discussing a regime in which
the quasar and Seyfert classifications indeed overlap. From their radio flux
it is clear that the Seyferts in our sample belong to the radio-quiet
class of AGN, yet they not only show jets, but the jets are also
kinematically important for the emission-line gas.  The jets we find
in Mrk~34 and Mrk~573 would in fact resemble some of the double
structures seen by Kellermann et al.~(1994) in radio-quiet quasars, if
placed at a larger distance and observed with lower sensitivity.

\section{Compact radio cores}
The advantage of the strong interaction in extended Seyfert jets for
the qualitative discussion of the importance of jets turns into a
major disadvantage if one tries to obtain quantitative statements
about jets, since any model naturally would require many
parameters. Hence, extended Seyfert jets do not serve well as standard
probes for the energetics of jets and one needs a different tool if
one wants to compare large samples of sources with each
other. Fortunately, jets in AGN are coherent structures with scales
from a few AU up to several megaparsecs and one can try to study
different parts of the jet in order to get the same answer.

For example, we can use the large, extended lobes of radio jets to
estimate their total power, e.g. by calculating their minimum energy
content from synchrotron theory or from their interaction with hot,
x-ray emitting gas and dividing by the life time of the sources
(e.g. derived from spectral aging). The derived powers (which are
often {\em lower} limits) are very high -- up to $10^{45-47}$ erg/sec
(Rawlings \& Saunders 1991) and the only reasonable place where such
enormous amounts of energy can be released is deep in the potential
well of a super-massive black hole; this is also indicated by VLBI and
variability observations (also gamma-ray observations), which suggest
that jets indeed come from a sub-parsec scale. However, ten
gravitational radii ($R_{\rm g}=GM_\bullet/c^2$) for an extremely
massive black hole of $M_\bullet=2\cdot10^9M_\odot$ correspond to
$3\cdot10^{15}$ cm, while the size of hotspots, where the jet
terminates, can be several kiloparsecs ($>3\cdot 10^{21}$cm;
e.g. Leahy et al.~1997) and the lobes are even larger. This yields an
expansion factor of $10^6$ and more. Hence, if the jets would suffer
adiabatic losses due to their lateral expansion (i.e. they would work
against the ISM all the way), their total energy losses would scale as
$r^{-2/3}$ for a relativistic plasma, and in our case we would have to
take losses of 4 orders of magnitude into account. That would require
the jets to start with initial powers of $10^{49-51}$ erg/sec,
corresponding to accretion rates of $10^{2-4}M_\odot$/yr and Eddington
luminosities for black holes with a mass of $10^{10-12}M_\odot$. From
all what we know today, this seems to be too high and one must
conclude, that, at least part of the way, the jet does not suffer
adiabatic losses and is not in pressure equilibrium with the ISM.

One region were this ``inflationary phase'' is likely to happen is
close to the nucleus, where flat spectrum radio cores are produced and
the energy density in powerful jets can be $1-100$ erg/cm$^3$ and
above, compared to $10^{-12}$ erg/cm$^{3}$ in the local ISM.  Hence,
here we will assume that indeed the jets are initially freely
expanding after they leave the nozzle and that the energy of the
rebounding shocks driven into the jet by the interaction with the ISM
is negligible compared to the overall power of the jet. This scenario
is in marked contrast to most numerical simulations of jets, where
usually a situation close to pressure-equilibrium is assumed. Hence,
we suppose that the jet will start to expand into a cone with the
opening angle of its Mach cone. A simple calculation shows that the
radio spectrum expected from such a configuration will be flat with a
constant brightness temperature as a function of size and frequency
(Blandford \& K\"onigl 1979). The flux of these radio cores is then
directly related to the jet power and the inclination angle, if
relativistic boosting is important, and environmental effects should
be of minor significance. Therefore, the compact radio core, much more
than any other part of the jet, allows us to directly probe
jet-related properties. Of course, in reality the situation will not be
quite as simple, e.g. the radial pressure gradient in the jet will
lead to some acceleration of the plasma and will tend to invert the
spectrum if the plasma is relativistic (Falcke 1996b) and at some point
external collimation and interaction with dense clouds will not be
completely negligible anymore.

Nevertheless, compact radio core fluxes have been used successfully to
investigate jet powers of large samples. For example in Falcke et
al. (1995) we compared the radio core fluxes of a sample of optically
selected quasars with model predictions of the flux distribution and
found that for radio loud quasars the total jet powers have to be of
similar magnitude as their accretion disk luminosities and their power
in the extended lobes. The same conclusion was reached by Celotti \&
Fabian (1993) and by Celotti et al.~(1997). Quite interestingly, the
latter paper used a large sample of quasars, but, in addition to a
simple radio core model, also used Synchrotron-Self-Compton (SSC)
theory to derive a jet power from radio flux and x-ray
emission. However, even though the basic result showed again the
equality between jet power and accretion disk luminosity, the
inclusion of the x-ray data seemed to increase the scatter
substantially rather than improving the earlier results, which were
based on the radio flux alone.

Despite differences in detail all results so far point to a connection
between the optical/UV emission and the radio core fluxes, thus
strengthening the idea of a jet/disk symbiosis. But, how universal is
this principle? How far does it go? Are there luminosities or source
types where this principle breaks down? What happens with jets in
quasars if the accretion rate becomes lower and lower? Will they die
completely, implying that accretion near the Eddington limit is
required for the jet formation, or will the jet just become
proportionally weaker, implying that jet formation is an integral part
of accretion physics? To learn more about this question one first has
to search for and then study jets in low-luminosity AGN.

\section{Compact radio cores in LINER galaxies}
Ho et al.~(1995 \& 1997) found that roughly one half of nearby
galaxies show signs of nuclear activity in the form of LINER or
Seyfert spectra. The bolometric luminosities of these AGN (excluding
the host galaxy of course) are in the range $10^{41}-10^{44}$
erg/sec. Heckman (1980) has speculated that LINER galaxies may
preferentially host compact radio cores in their nuclei; according to
Falcke (1996b\&c) these cores can be interpreted as scaled down
versions of the compact radio cores and jets in radio-loud quasars.

To test this, we have performed a VLA A-array survey at 2cm of 48
nearby LINER galaxies (Falcke, Wilson, \& Ho, in prep.) from the Ho et
al.~(1995) sample, to search for compact, flat-spectrum radio
nuclei. The $5\sigma$ detection limit of the survey was $1$ mJy. In
total we detected 21 galaxies at this wavelength. Twelve of them have
flat and nine have steep spectra. The spectral indices were derived
from a comparison with C and X-band observations of the same sample by
Van Dyk \& Ho (1998) and from additional literature data. We note that
out of the 12 flat-spectrum sources, 10 are in spiral galaxies.

Our detection rate of flat-spectrum, compact nuclei at 2cm is
relatively high---especially in the spiral galaxies---and confirms the
initial hunch that LINERs would make a good sample to detect compact
radio nuclei. For comparison, Vila et al.~(1990) looked at a sample of
Sbc galaxies with nuclear radio components and only detected 2
flat-spectrum nuclei in a sample of 27 galaxies---both 
were LINERs. In elliptical galaxies, however, the detection
rate of compact nuclei is higher (Wrobel \& Heeschen 1991).

The mean spectral index we find for our flat spectrum sources is
$\alpha=+0.15$ which, however, is rather meaningless since it is based
on non-simultaneous data. Comparison with earlier observations shows
that some of our sources have varied in flux by up to a factor of
five.

\begin{figure}
\centerline{
\psfig{figure=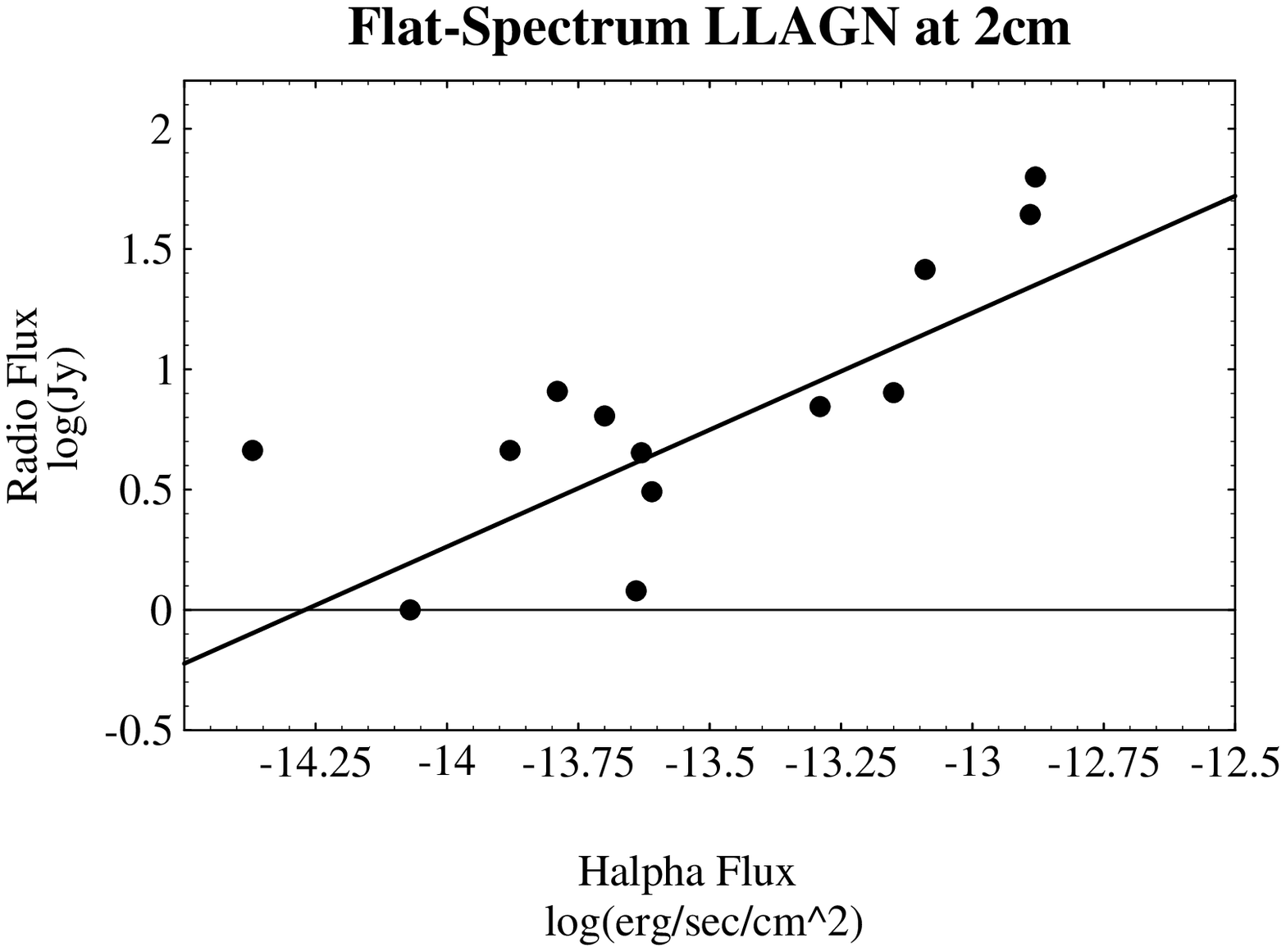,width=0.48\textwidth,bbllx=2.7cm,bblly=7.8cm,bburx=19cm,bbury=20.2cm}
\psfig{figure=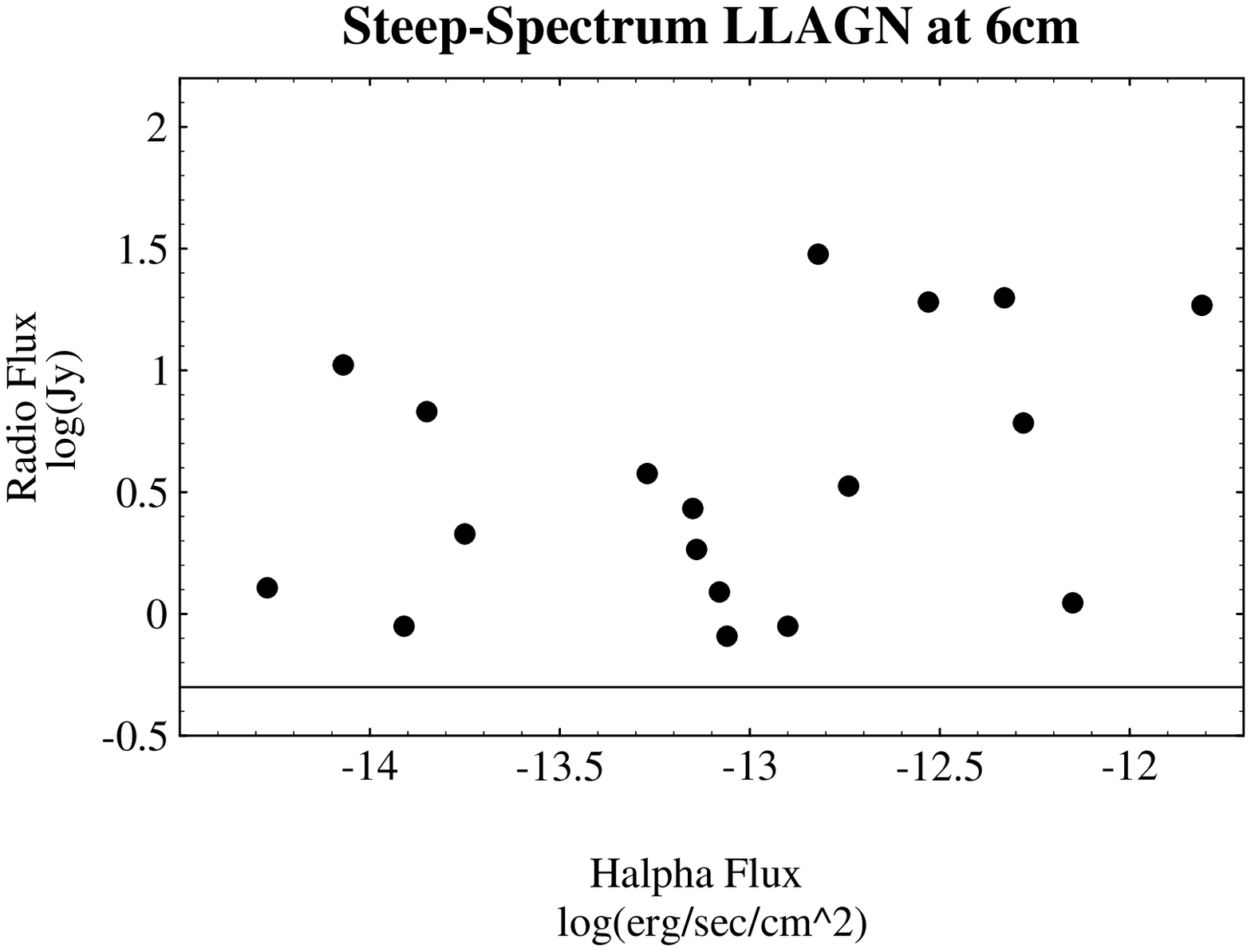,width=0.48\textwidth,bbllx=2.7cm,bblly=7.8cm,bburx=19cm,bbury=20.2cm}
}
\caption[]{Plots of radio fluxes versus nuclear H$\alpha$ fluxes for a sample
of nearby LINER galaxies. Left: 2cm emission of LINERS in the sample 
with a compact, flat spectrum emission; Right: 6cm emission of LINERS 
in the sample with steep spectrum emission.
}
\end{figure}

Of course, the mere fact that we find these radio nuclei in LINERs
does not prove yet that these radio cores are indeed related to the
active nucleus or that they are jets. We therefore looked at the
relation between radio and optical H$\alpha$ flux in our galaxies
and found a reasonably good correlation for all galaxies with a flat
spectrum (Fig.~5, left panel). This is in line with earlier claims of
a connection between optical and radio activity (Ekers \& Ekers 1973,
O'Connell \& Dressel 1978). The significance of this possible
correlation is greatly strengthened if one plots the same diagram for
the LINER galaxies with steep spectrum emission (Fig.~5, right panel)
where there is not even a hint of a correlation.

\begin{figure}
\centerline{\psfig{figure=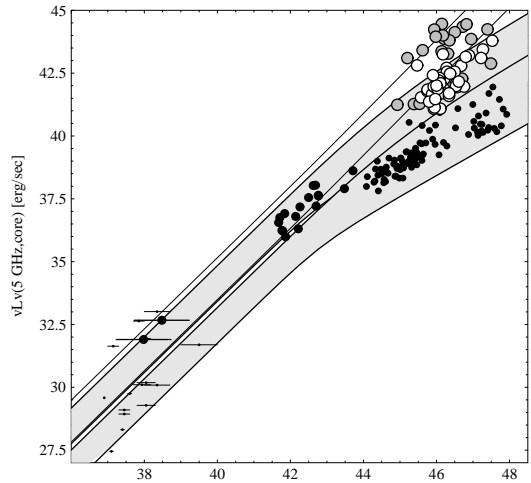,width=0.5\textwidth,bbllx=2.5cm,bblly=6.3cm,bburx=19.1cm,bbury=21.7cm}}
\caption[]{Correlation between accretion disk luminosity (i.e. nuclear
optical+UV luminosity for AGN) and monochromatic radio core luminosity
at 5 GHz. The shaded bands are the theoretical predictions as
presented in Falcke \& Biermann (1996) for radio core luminosities as a
function of accretion disk luminosity for relativistic jets with
randomly oriented inclination angles. The radio cores of the newly
added LINER galaxies are given by big dots, quasars are given by open
(steep-spectrum, radio-loud) and filled (flat-spectrum,
radio-loud) circles, as well as smaller dots above $L_{\rm
disk}=10^{44}$ erg/sec (radio-quiet). Sources in the lower left are Sgr A*, M31* and stellar
mass black holes (see Falcke \& Biermann 1996 for more details).}
\end{figure}

Therefore, we conclude that the radio cores in LINERs are indeed part
of the central engine. Moreover, we can compare the radio and
emission-line luminosities with the jet/disk model by Falcke \&
Biermann (1996), to learn more about their nature.  The model
predicted a specific radio/nuclear luminosity correlation for
low-power AGN and is based on the assumption that accretion disk
luminosity and jet power in AGN are coupled by a universal constant of
order unity. We note that this may remain true even for advection
dominated disks, as have been discussed for LINERs (Lasota et
al. 1996), if the radiative efficiency of the radio and the optical
emission in a jet/disk system are reduced in the same way (i.e. the
jet power is proportional to the energy dissipated in the disk rather
than to the accretion rate).

For a randomly selected (and randomly oriented) sample, the width
(`scatter') of the radio-to-nuclear UV distribution is given by the
typical Lorentz factor of the jets. In Fig.~6 we reproduce (without
changing any parameters) the figure from Falcke \& Biermann (1996),
where the model predictions for low-power AGN was given as shaded
bands. We then converted the {\it narrow} H$\alpha$ line-luminosities
of the LINERs with detected flat-spectrum nuclei to optical+UV
luminosities, using the same proportionality factors as for the
quasars\footnote{See Falcke et al.~(1995) \& Falcke (1996a). The exact
conversion factors for LINERs require of course a more thorough
discussion. For a few examples (e.g. M81, NGC 4252) this method at
least seems to give a reasonable estimate for the nuclear luminosity},
and we inserted them into the correlation (big dots).  Obviously, the
LINERs fall exactly into the range predicted for low-luminosity, {\it
radio-loud} jets. This confirms a preliminary version of this diagram
which was presented in Falcke (1996c), but was based only on a few
ill-selected, famous LINER galaxies.

The result not only strongly suggests that LINERs do have powerful
nuclear radio jets---for some individual cases this was known already
(e.g. M87; NGC4258, Herrnstein et al.~1997; M81, Bietenholz et
al. 1996, etc.)---but is also consistent with mildly relativistic
Lorentz factors around $\gamma_{\rm j}\simeq2$ as used in the
model. That should be compared with Lorentz factors of
$\gamma\simeq6-10$ derived with the same method for radio-loud quasars
(Falcke et al.~1995).  For the lower Lorentz factor in LINERs (and
also in the Galactic superluminal sources) one can give a very simple
explanation, since this terminal velocity is naturally obtained by a
relativistic plasma in a simple pressure driven jet (Falcke 1996b). To
explain the velocities in radio-loud and radio-quiet quasars, however,
one needs an extra mechanism that provides the additional push
necessary to go beyond $\gamma_{\rm j}=3$.

\section{Stellar Mass Black Holes}
In the sequence of this paper we have slowly moved towards lower and
lower luminosities, hence we cannot conclude this paper without a
brief discussion of stellar mass black holes, which in terms of
absolute luminosity are at the bottom of this distribution of black hole,
jet, \& disk systems.  After all, the discovery of the microquasars
has almost led to a revolution in the study of jets. Mirabel et
al.~(1992) discovered that an X-ray binary, 1E1740.7--2942, near the
Galactic Center had symmetric radio lobes not very different from
FR\,I radio galaxies -- the major difference only being that this jet
was many orders of magnitude smaller and less powerful than those
found in AGN. Later, Mirabel \& Rodriguez (1994) also discovered a
compact radio jet in GRS~1915+105 with apparent superluminal motions,
clearly showing that also in the jets of stellar mass black holes
relativistic speeds are obtained. The existence of relativistic jets
associated with stellar mass black holes was already predicted by
Hjellming \& Johnston (1988) and Falcke (1994) from the analogy to
extragalactic systems.  The big advantage of these "small" systems
compared to AGN is that their time scale is accordingly shorter. The
study of the variability of GRS~1915+105 in x-rays and radio has led to
a number of very interesting results, especially concerning the
connection between the accretion process and the formation of a
jet. For example, in a recent paper Mirabel et al.~(1998) found an
intriguing correlation between radio outbursts and x-ray flares: it
seemed that whenever there was a radio outburst, there was a sudden
drop in x-ray emission. Because the x-ray emission is probably coming
from the inner parts of an accretion disk, Mirabel et al.~have
suggested that the emission of radio blobs is caused by ejection of
material from the inner edge of the accretion disk leading to its
temporary disappearance. In Falcke \& Biermann (1995) we had already
argued for energetical reasons that a large fraction of the energy
produced in the inner disk is not radiated away but ``dissipated''
into the jet.  If indeed correct---and further studies will
eventually give a much clearer picture of these "laboratory jet
systems", this would finally confirm the jet/disk symbiosis picture
and will even allow us to uncover some of the details of the coupling
between jet and disk.

Another argument for the validity of the jet/disk-symbiosis model comes
from the direct comparison of the observed jet parameters and the
model predictions, since the basic jet (and disk) parameters for
GRS~1915+105 are probably more accurately determined than for any other
source and are extremely constraining.  Like in the previous section
we will again concentrate on the radio cores, where the jet/disk
symbiosis model predicts sizes and fluxes of the core as a function of
frequency and accretion disk luminosity. The most recent version was
published in Falcke (1996b) applied to the radio nucleus M81* in the
LINER galaxy M81. The critical difference between the Falcke (1996b)
model and the earlier version is just that the velocity field of the
jet is calculated self-consistently, taking the acceleration of the
jet plasma due to the pressure gradient along the jet (and accordingly
the adiabatic cooling) into account and thus reducing the number of
free parameters. The model simply assumes that gas is `boiled' somehow
to its maximal temperature, thus producing a fully relativistic plasma
which leaves a nozzle at $\sim10$ (the exact value is not critical)
gravitational radii ($R_{\rm g}=GM_\bullet/c^2$) and then freely
expands into the vacuum, yielding velocities given by Eq.~2 of Falcke
(1996b). The `radio loud' model also assumes a `maximal' equipartition
jet which is the most efficient radio emitter and generally fixes the
main parameters (i.e. $\mu_{\rm p/e}$, $x_{\rm e}$, $q_{\rm j/l}$, see
below and Falcke 1996b for details).

We can now use the pressure-driven jet model for M81* without any
further modifications and Eqs.~9 and 10-12 of Falcke (1996b) can be
directly applied to GRS~1915+105 without any fiddling of
parameters. Here we have rederived the equations for a distance of
12.5 kpc, and an inclination angle of $i=70^\circ$ and $i=110^\circ$
(for jet and counter jet) as derived for GRS~1915+105 by Mirabel \&
Rodriguez (1994). Moreover, for simplicity  we only give the
exponents rounded to one significant digit after the decimal, which
allows one to evaluate the sums easily. Finally, we have fixed the
intrinsic parameters of the model---which should be of order unity for
a radio loud (maximal) jet, as mentioned above---at the canonical
values used in Falcke (1996b) and Falcke \& Biermann (1995),
i.e. $\mu_{\rm p/e}=1.5$, $x_{\rm e}=0.5$, and $q_{j/l}=0.5$. The
first two parameters define that the typical Lorentz factor of the
radiating electrons are of the order $10^2-10^3$ (here $\gamma_{\rm
e}=450$), and the last parameter states that the amount of energy
radiated by the accretion disk is equal to the total energy input into
the jet. The equality of jet and disk power as well as the constancy
of these intrinsic parameters is one of the basic principles of the
jet/disk-symbiosis, allowing to scale the model over many orders
magnitude and still making very specific predictions. The model in
this simplified version then predicts the radio emission of the core
to be

\begin{equation}
S_{\nu}=20\,{\rm mJy}\cdot 
\left(L_{disk}\over10^{39}\,{erg/sec}\right)^{1.4}
\left(M_\bullet\over33M_{\sun}\right)^{0.2}
\left(\nu\over8.5\,{\rm GHz}\right)^{0.2},
\end{equation}
for an accretion disk luminosity $L_{\rm disk}$ (here equal to the jet
power!), a black hole mass $M_\bullet$, and the observing frequency
$\nu$. The expected size of the core (taking jet and counter jet into
account) is\footnote{Please note that
the similar Eq.~6 in Falcke \& Biermann (1996) has the wrong sign
in both exponents, the scaling there has to go as
$z\propto\nu^{-1}L^{-2/3}$ -- the rest of the paper is not affected
by this typo.}

\begin{equation}
z({\nu})=6\,{\rm mas}\cdot 
\left(L_{disk}\over10^{39}\,{erg/sec}\right)^{0.4}
\left(M_\bullet\over33M_{\sun}\right)^{0.1}
\left(\nu\over8.5\,{\rm GHz}\right)^{-0.9}.
\end{equation}

The parameters $L_{\rm disk}$ and $M_{\bullet}$ are scaled to the
appropriate values for GRS~1915+105 currently discussed in the
literature (e.g.~Mirabel et al.~1997; Morgan et al.~1997). Of course,
this stationary model is only suitable for the quiescent but not the
outburst phase. The predicted values fit very well the VLBA
observations by Mirabel et al.~(1998), where they give a quiescence
source size of $\sim6$ mas (major axis) and fluxes in the range 20-100
mJy. Moreover, the expected scaling of the core size
($\propto\nu^{-0.9}$) is very close to the observed one
($\propto\nu^{-1}$) and the model also predicts a flat average
spectral index for the core of $\alpha=+0.2$. The velocity of the jet
in the model grows asymptotically as determined by Eq.~2 in Falcke
(1996b), yielding $\beta=$0.92 at $10^4 R_{\rm g}$ and $\beta=0.96$ at
the scale of a few mas ($10^8R_{\rm g}$), where the radio emission is
coming from, in good agreement with the observed values. Clearly, the
pressure gradient effect must be at work at least to some degree here,
since Mirabel \& Rodriguez (1994) found that indeed the blobs expand
with $0.2c$ at larger scales, thus finding direct evidence that the
bulk of the plasma can indeed be described as an expanding
relativistic gas. From the calculated values above one can also see
that the pressure gradient alone is already sufficient to explain the
jet velocities inferred from the observations. Consequently, in
GRS~1915+105 there is no need for any additional bulk acceleration
mechanisms, e.g. as in centrifugally driven MHD jet models. Finally,
one can turn our energy argument around: since we have used the most
efficient type of radio jet and assumed that $L_{\rm disk}=Q_{\rm
jet}$ (the total jet power), we can say that under any circumstances
$Q_{\rm jet}\ga10^{39}$ is a good lower limit for the total jet power
in GRS~1915+105---irrespective of ones belief in the jet/disk
coupling.

To conclude this section, it should be noted that the simple idea of a
jet/disk coupling, parametrized in a very simple form is able to
describe objects from quasars to microquasars, which are separated in
luminosity by eight orders of magnitude, to such a detail and
accuracy. The same model can describe the radio core in M81 as well as
in GRS~1915+105 equally well, by just changing the accretion rate and
the inclination angle. We have argued for the existence of this
scaling for a number of years now and with more and more data the
statement becomes stronger and stronger, so that one can be confident
that there is something fundamental behind it.

\section{Conclusions}
Today we can make at least one firm statement, namely that there is no
known class of active sources, where one suspects an accreting black
hole as the central engine, where jets cannot be found in at least a
few members of this class. Consequently {\it jet formation must be in
principle possible under almost all circumstances}, whether there are
huge or small luminosities, huge or small Eddington-luminosities, or
huge or small sizes, there always seems to be a way to make it
possible. Given the many observed stellar (Herbig-Haro) jets, this may
even be true if the central object is not a black hole.  

The question remains, however, whether jets always form in each and
every case, i.e. whether each accreting black hole always produces a
jet? Even though I suspect the answer to be a strong "yes", with only
a feeble "but", a conclusive answer cannot be given yet, since in many
cases our sensitivity is not good enough to securely prove or exclude
the existence of jets. Much of the progress in the detection of
astrophysical jets in recent years has made use of the superior
resolution and sensitivity of the VLA, often pushing its ability to
the limits, so that further substantial progress may require a further
step in radio technology. For Seyferts, the combination of high
resolution radio maps with observations of the narrow emission line
regions presented in a number of papers, including this one, has now
confirmed earlier claims that outflows are a crucial ingredient to the
study of Seyferts and therefore most likely also to AGN physics in
general. Still, there is a whole universe in front of us that wants to
be explored, as for example the low-luminosity AGN in our neighborhood
which are receiving more and more attention in recent years. One
result is already clear: jets are here to stay, and their importance
should not be underestimated as perhaps done in the past by many
astronomers not concerned with radio astronomy.

\bigskip\noindent
{\it Acknowledgment:} This work was done in collaboration with a number of colleagues who I need to thank for their cooperation, especially A.S. Wilson, C. Simpson, A. Patnaik, W. Sherwood, \&  L.C. Ho, and P.L. Biermann.
This summary paper is based in some parts on material reported in Falcke et al.~(1998a\&b). The research was supported by NASA under grants NAGW-3268, NAGW4700, NAG8-1027, and by the DFG, grant Fa 358/1-1\&2.

\end{document}